\documentclass{article}

\usepackage{arxivsimplified}

\usepackage[utf8]{inputenc} 
\usepackage[T1]{fontenc}    
\usepackage{hyperref}       
\usepackage{url}            
\usepackage{booktabs}       
\usepackage{amsfonts}       
\usepackage{nicefrac}       
\usepackage{microtype}      
\usepackage{lipsum}

\usepackage{graphicx}
\graphicspath{{figs_pdf/}}
\usepackage{subfigure}
\usepackage{cite}
\usepackage[dvipsnames]{xcolor}
\usepackage{amsmath}
\usepackage{blindtext}
\usepackage{amsfonts}
\usepackage{multirow}
\usepackage{array}
\usepackage{mathtools}
\usepackage{listings}
\usepackage{xcolor}
\definecolor{codegreen}{rgb}{0,0.6,0}
\definecolor{codegray}{rgb}{0.5,0.5,0.5}
\definecolor{codepurple}{rgb}{0.58,0,0.82}
\definecolor{backcolour}{rgb}{0.95,0.95,0.92}
\usepackage{color}
\usepackage{xcolor}
\definecolor{rev1}{rgb}{0,0,0}
\usepackage{esvect}
\usepackage{tabularx}
\newcolumntype{L}[1]{>{\raggedright\arraybackslash}p{#1}}
\newcolumntype{C}[1]{>{\centering\arraybackslash}p{#1}}
\newcolumntype{R}[1]{>{\raggedleft\arraybackslash}p{#1}}

\DeclareMathOperator{\argmin}{arg\,min} 

\usepackage{algorithm}
\usepackage{algorithmic}
\usepackage{enumitem}
\setlist[itemize]{leftmargin=*}

\lstdefinestyle{mystyle}{
    backgroundcolor=\color{backcolour},   
    commentstyle=\color{codegreen},
    keywordstyle=\color{magenta},
    numberstyle=\tiny\color{codegray},
    stringstyle=\color{codepurple},
    basicstyle=\ttfamily\footnotesize,
    breakatwhitespace=false,         
    breaklines=true,                 
    captionpos=b,                    
    keepspaces=true,                 
    numbers=left,                    
    numbersep=5pt,                  
    showspaces=false,                
    showstringspaces=false,
    showtabs=false,                  
    tabsize=2
}
 
\usepackage{scalerel,stackengine}
\stackMath
\newcommand\reallywidehat[1]{%
\savestack{\tmpbox}{\stretchto{%
  \scaleto{%
    \scalerel*[\widthof{\ensuremath{#1}}]{\kern-.6pt\bigwedge\kern-.6pt}%
    {\rule[-\textheight/2]{1ex}{\textheight}}
  }{\textheight}%
}{0.5ex}}%
\stackon[1pt]{#1}{\tmpbox}%
}

\title{Multi-fidelity information fusion with concatenated neural networks}

\author{
    Suraj Pawar  \\
    School of Mechanical \& Aerospace Engineering,\\
    Oklahoma State University, \\
    Stillwater, Oklahoma - 74078, USA.\\
    \texttt{supawar@okstate.edu} \\
    \And
    Omer San \\
    School of Mechanical \& Aerospace Engineering,\\
    Oklahoma State University, \\
    Stillwater, Oklahoma - 74078, USA.\\
    \texttt{osan@okstate.edu} \\
    \And
    Prakash Vedula \\
    School of Aerospace \& Mechanical Engineering,\\
    The University of Oklahoma, \\
    Norman, OK 73019, USA. \\
    \And
    Adil Rasheed \\
	Department of Engineering Cybernetics, \\
	Norwegian University of Science and Technology, \\
	7465 Trondheim, Norway. \\
    Department of Mathematics and Cybernetics, \\
    SINTEF Digital, \\
    7034 Trondheim, Norway. \\
    \AND
    Trond Kvamsdal \\
    Department of Mathematics and Cybernetics, \\
    SINTEF Digital, \\
    7034 Trondheim, Norway. \\
    Department of Mathematical Sciences, \\
    Norwegian University of Science and Technology, \\
    7491 Trondheim, Norway.
}

\begin{document}
\maketitle

\begin{abstract}
Recently, computational modeling has shifted towards the use of statistical inference, deep learning, and other data-driven modeling frameworks. Although this shift in modeling holds promise in many applications like design optimization and real-time control by lowering the computational burden, training deep learning models need a huge amount of data. This big data is not always available for scientific problems and leads to poorly generalizable data-driven models. This gap can be furnished by leveraging information from physics-based models. Exploiting prior knowledge about the problem at hand, this study puts forth a concatenated neural network approach to build more tailored, effective, and efficient machine learning models. For our analysis, without losing its generalizability and modularity, we focus on the development of predictive models for laminar and turbulent boundary layer flows. In particular, we combine the self-similarity solution and power-law velocity profile (low-fidelity models) with the noisy data obtained either from experiments or computational fluid dynamics simulations (high-fidelity models) through a concatenated neural network. We illustrate how the knowledge from these simplified models results in reducing uncertainties associated with deep learning models applied to boundary layer flow prediction problems. The proposed multi-fidelity information fusion framework produces physically consistent models that attempt to achieve better generalization than data-driven models obtained purely based on data. While we demonstrate our framework for a problem relevant to fluid mechanics, its workflow and principles can be adopted for many scientific problems where empirical, analytical, or simplified models are prevalent. In line with grand demands in novel physics-guided machine learning principles, this work builds a bridge between extensive physics-based theories and data-driven modeling paradigms and paves the way for using hybrid physics and machine learning modeling approaches for next-generation digital twin technologies.    
\end{abstract}


\section{Introduction}
The modeling of spatiotemporal dynamics of multiscale and multiphysics systems is an open problem relevant to many scientific and engineering applications. For instance, wind energy is a highly complex system whose dynamics is governed by global atmospheric processes to turbulent boundary layer formed around the blades that span nine orders of magnitudes \cite{veers2019grand}. Over the past several decades, we have improved our understanding of such multiphysics systems by developing accurate numerical models for governing equations of the system, such as Navier-Stokes equations for fluid flows. However, these numerical models can be computationally prohibitive, especially for nonlinear multiscale systems, and their use in real-time optimization and control is scarce. The recent advancement in machine learning (ML) and deep learning (DL) holds the great potential for tackling the challenge of modeling and analysis of high-dimensional systems and has been successful in diverse applications, such as fluid mechanics \cite{brunton2020machine}, earth science \cite{reichstein2019deep}, and material science \cite{schmidt2019recent}. These advances have been driven by a vast amount of data generated from high-fidelity numerical simulations, experimental and satellite measurements, and computing power along with the emergence of effective and efficient algorithms that can extract relevant patterns from the data. ML/DL techniques has been successfully applied for turbulence closure modeling \cite{duraisamy2019turbulence}, super-resolution of climate data \cite{stengel2020adversarial}, predicting clustered weather patterns \cite{chattopadhyay2020predicting}, reduced-order modeling \cite{fresca2021comprehensive}, and many more.   

ML/DL models are capable of providing insights from data, exploiting these insights in building predictive tools, and continuously updating themselves as the new streams of data get available. Despite these advantages, ML/DL techniques lack interpretability and suffer from the curse of dimensionality. The interpretability issue can be addressed by understanding the physical implications of ML/DL models \cite{mcgovern2019making}, and understanding the neural network correlations discovered from the data \cite{montavon2019layer, ebert2020evaluation}. By the curse of dimensionality, we mean that DL models are data-hungry in nature. For instance, Bonavita and
Laloyaux \cite{bonavita2020machine} showed that the amount of the training data for nonlinear dynamical systems grows exponentially with the dimensionality of the system. Furthermore, pure ML/DL models lead to poor extrapolation/generalization, i.e., they fit the observations data very well, but predictions may be poor and physically inconsistent for data beyond the distribution of the training dataset. To this end, physics-informed learning algorithms that leverage prior knowledge based on the physical and mathematical understanding of the system are proposed in several studies \cite{karniadakis2021physics}. 

\begin{figure*}[ht]
\centering
\includegraphics[width=0.99\textwidth]{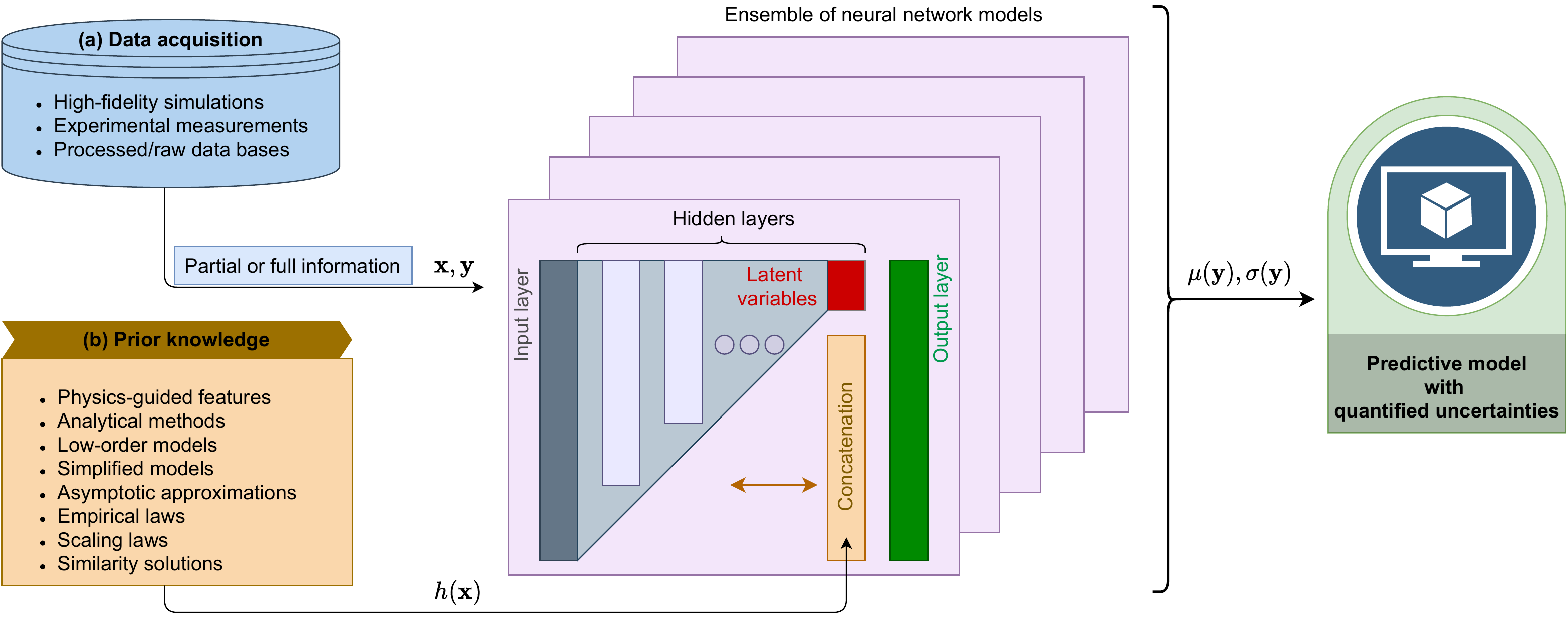}
\caption{The proposed multi-fidelity information fusion framework. The prediction from the low-fidelity models (e.g., self-similarity solution or law of the wall) is concatenated with the latent variables at the certain hidden layer of the neural network. An ensemble of neural networks is trained using the negative log-likelihood loss function to estimate the uncertainty associated with the prediction. Here, $\mathbf{x}$ refers to independent variables or design parameters, or $\mathbf{y}$ indicates the quantities of interest, and $h(\mathbf{x})$ represents a low-fidelity or simplified model that is fast-to-compute.  
}
\label{fig:pgml}
\end{figure*}


There are two main techniques based on inductive biases, and learning biases to embed physics into ML/DL models in combination with the observations data \cite{karniadakis2021physics}. Inductive biases techniques relate to building tailored ML/DL model architecture that exploits the prior knowledge about the problem at hand to build physically consistent data-driven models. The representative examples includes embedding invariance property into neural network architecture \cite{ling2016reynolds,zanna2020data}, imposing conservation laws of physical quantities or analytical constraints into neural network \cite{mohan2020embedding,beucler2021enforcing,greydanus2019hamiltonian}, using prediction from simplified models as the bias \cite{pawar2021physics}, and equivariant transformer networks \cite{tai2019equivariant}. The other approach based on learning biases imposes constraints such as governing equations in a soft manner by penalizing the loss function of ML/DL models. Some of the examples of this approach pertain to physics-informed neural networks \cite{raissi2019physics}, statistically constrained generative adversarial networks \cite{wu2020enforcing}, and Bayesian framework with auto-regressive model \cite{geneva2020modeling}. There is also a class of hybrid analysis and modeling approaches that utilizes pure data-driven and physics-based models in tandem \cite{san2021hybrid}. While many methods have been demonstrated to be successful in enforcing physics into ML/DL models, they offer many possibilities to fuse domain knowledge to improve the generalizability and data efficiency of data-driven models \cite{kashinath2021physics}.

In this work, we introduce a tailored neural network architecture based on a concatenation layer for surrogate modeling of high-fidelity data given the solution from a computationally inexpensive low-fidelity model. There is a hierarchy of numerical models ranging from simple empirical relations to highly accurate numerical discretization-based models across all scientific disciplines. For example, flow around airfoils can be modeled using tools extending from panel methods \cite{hess1990panel} to the fully resolved direct numerical simulation (DNS) \cite{moin1998direct}. Another example is the wake modeling of wind turbines, where fast but typically inaccurate analytical models are adopted for tasks like layout optimization and wind farm control \cite{archer2018review}. However these models are insufficient to take the unsteady nature of interactions of turbine wake with other wakes as well as atmospheric turbulence into account, and such effects can be modeled with computationally demanding but accurate models like large-eddy simulation (LES) \cite{breton2017survey}. Our work draws inspiration from these multi-fidelity modeling approaches and exploits the real-time prediction from low-fidelity models to inform a DL model of high-fidelity observations. The information fusion from the multi-fidelity sources of data leads to a robust and generalizable surrogate model in comparison to purely data-driven model trained on high-fidelity data. Figure~\ref{fig:pgml} graphically illustrates the proposed data fusion approach from multimodal data streams in the process of generating physics-guided machine learning (PGML) models.    

In a nutshell, our report puts forth a novel data-driven framework to take prior knowledge about the system into account when generating a black-box deep learning predictive model. How should we inject physics and domain knowledge into machine learning models? How deep learning can be constructed as a trustworthy approach toward more accurate real-time prediction of nonlinear complex systems such as turbulent flows? In this paper, these are the fundamental research questions that we tackle, and provide our insights.

We demonstrate the application of our framework for boundary layer flows. Boundary layer phenomenon is one of the most important flows and is of engineering concern in many scientific and industrial applications \cite{smits2013wall}. The behavior of flow in the boundary layer has implications on the drag force in ship hulls and aircraft, the energy required to move oil through pipes, and the distribution of heat in the atmosphere, and therefore boundary layer flows are extensively studied in the literature \cite{jimenez2012cascades, bhaganagar2004effect, wu2009direct}. Given that boundary layer flows are prevalent in engineering applications, building a computationally efficient and accurate surrogate model is of paramount importance for online tasks like boundary layer control to achieve lift enhancement, noise mitigation, drag reduction, and wall cooling. Additionally, boundary layer flows can be described using models that have different levels of fidelity spanning analytical models to direct numerical simulations and hence represent an interesting test case for illustrating the effectiveness of concatenated neural networks. As we will detail in Methods section, our multi-fidelity data-fusion framework approach consists of a concatenated neural network architecture where the self-similarity solution is a low-fidelity model and the Reynolds-Averaged Navier-Stokes Equations (RANSE) solver is a high-fidelity model. This framework can be easily scaled to large problems such as wake behind bluff bodies with information from many models fused to predict the high-fidelity data. 
While in this work we consider only two levels of fidelity, the proposed framework can be applied for blending information from various levels of fidelity. Moreover, the neural architecture search tools can be utilized to discover more complex and optimal architectures automatically \cite{elsken2019neural}.





\section{Results}
In this section, we demonstrate the capability of the proposed approach to reconstruct laminar and turbulent boundary layer flows around the flat plate. 

\begin{figure*}[t!]
\centering
\includegraphics[width=0.99\textwidth]{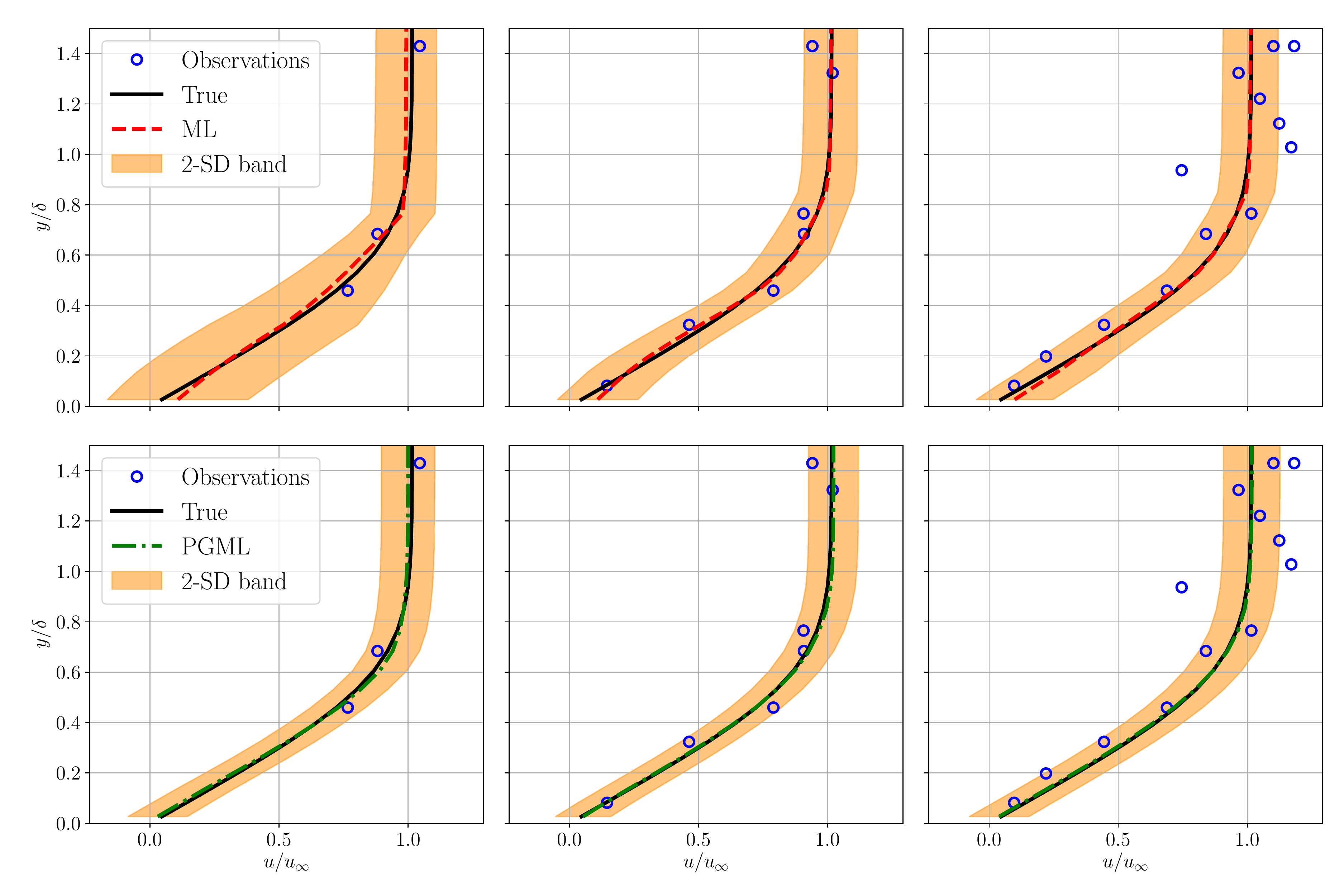}
\caption{Boundary layer prediction for laminar flat plate flow at $x/L=0.5$ along with the observations used for training the ML and PGML model. The amount of observations data used for training the model is 10\% (left), 30\% (middle), and 50\% (right). The shaded area corresponds to two standard deviation (2-SD) band.}
\label{fig:laminar}
\end{figure*}

\subsection{Laminar flow past a flat plate}
We refer to a simple feed-forward neural network as the machine learning (ML) model and the concatenated neural network is called the physics-guided machine learning (PGML) model. The ML model is trained purely based on the data, while the PGML model uses the prediction from physics-based low-fidelity model (Blasius equations, see Equation~\ref{eq:blasius} and \ref{eq:blasius2}) along with the data. Figure~\ref{fig:laminar} shows the profile of horizontal component of velocity versus distance from the wall, at the location $x/L = 0.5$, for different amounts of data used for training the ML and PGML models. The velocity is normalized with the freestream velocity and the vertical distance is normalized using the boundary layer thickness for laminar flow over the flat plate. The ML model fails to capture the accurate velocity profile when the velocity field information at 10\% of locations within the computational domain is utilized for training. The mean velocity profile predicted by the PGML model is highly accurate with just 10\% of the observations. The predicted mean velocity profile is also accompanied by a confidence interval spanning one standard deviation (SD) on either side of the mean velocity profile and is an outcome of the uncertainty quantification mechanism built into deep ensembles. Deep ensembles achieve uncertainty quantification by training an ensemble of neural networks with the negative log-likelihood loss function. The uncertainty estimate associated with the PGML model is lower than the ML model and this is particularly notable near the wall within the boundary layer, i.e., for $y/\delta < 0.8$.

\subsection{Turbulent flow past a flat plate}
\begin{figure*}[htbp]
\centering
\includegraphics[width=0.99\textwidth]{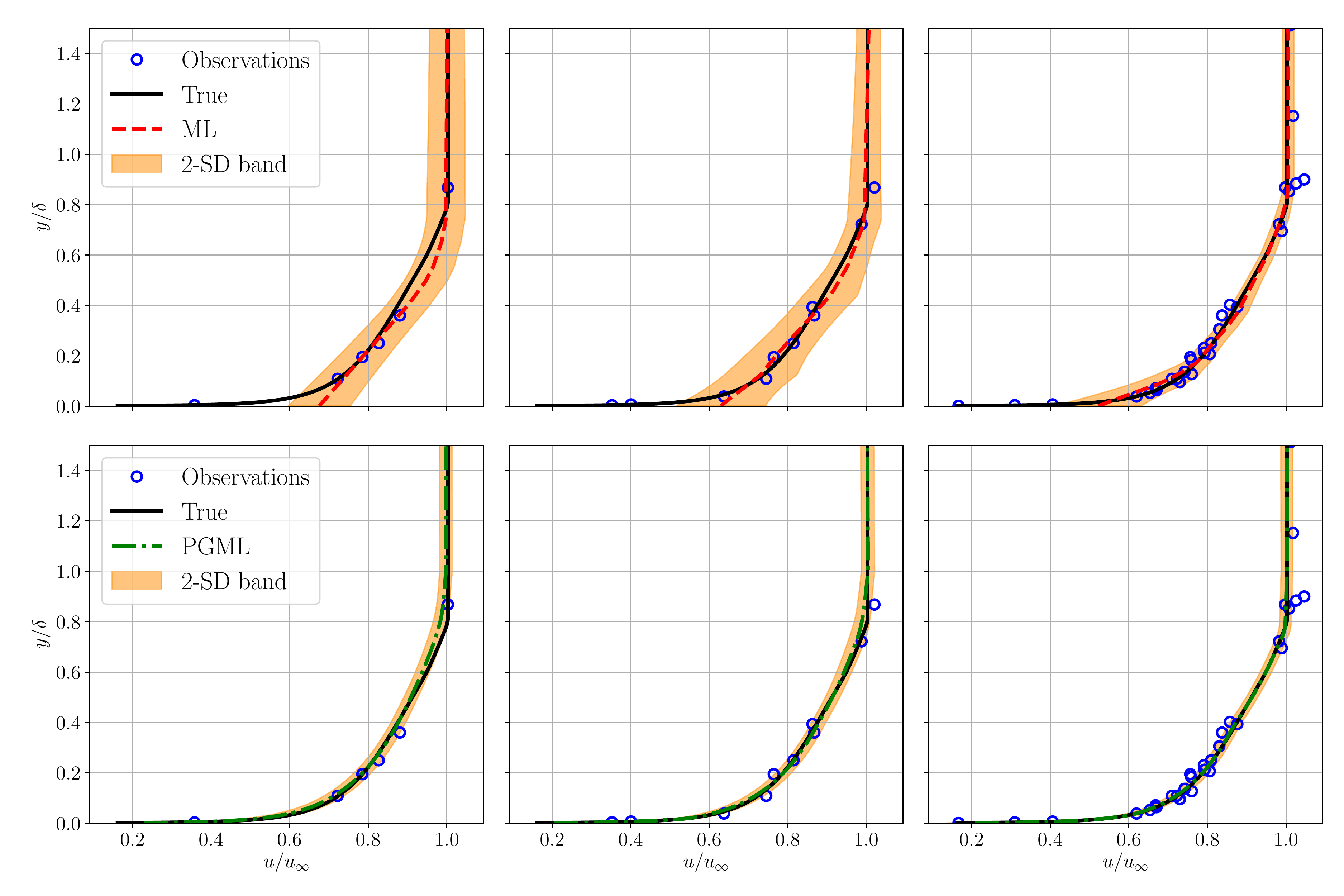}
\caption{Boundary layer prediction for turbulent flat plate flow at $x/L=0.5$ along with the observations used for training the ML and PGML model. The amount of observations data used for training the model is 5\% (left), 10\% (middle), and 30\% (right). The shaded area corresponds to two standard deviation (2-SD) band.}
\label{fig:turbulent}
\end{figure*}

Next, we evaluate the performance of the proposed PGML approach for reconstruction of turbulent boundary layer flow over a (smooth) flat plate. The prior knowledge we concatenate in this case is the one-seventh power law velocity profile (i.e., see Equation~\ref{eq:power_law}). Figure~\ref{fig:turbulent} displays the variation of the normalized velocity profile in the vertical direction at $x/L=0.5$ for different amounts of data used for training the ML and PGML models. We can observe that the ML model performs very poorly for the data-sparse regime (i.e., 5\% and 10\% of the observations). Such situations are very common in scientific applications where a collection of high-fidelity data either from experiments or numerical simulations can be prohibitive. The PGML model on the other hand leads to an accurate prediction by exploiting the correlation between low- and high-fidelity data. The prediction from the ML model is also not reliable as indicated by the high width of the confidence band. The slope of the boundary layer profile $\partial u / \partial y$ at the wall determines the skin friction drag along the wall. This quantity is not predicted accurately with the ML model and this can lead to poor estimation of the quantity of interests like the total drag. The PGML model is successful in predicting the correct slope of the velocity profile at the wall and therefore will lead to a more accurate estimation of total drag. 

\begin{figure*}[htbp]
\centering
\includegraphics[width=0.99\textwidth]{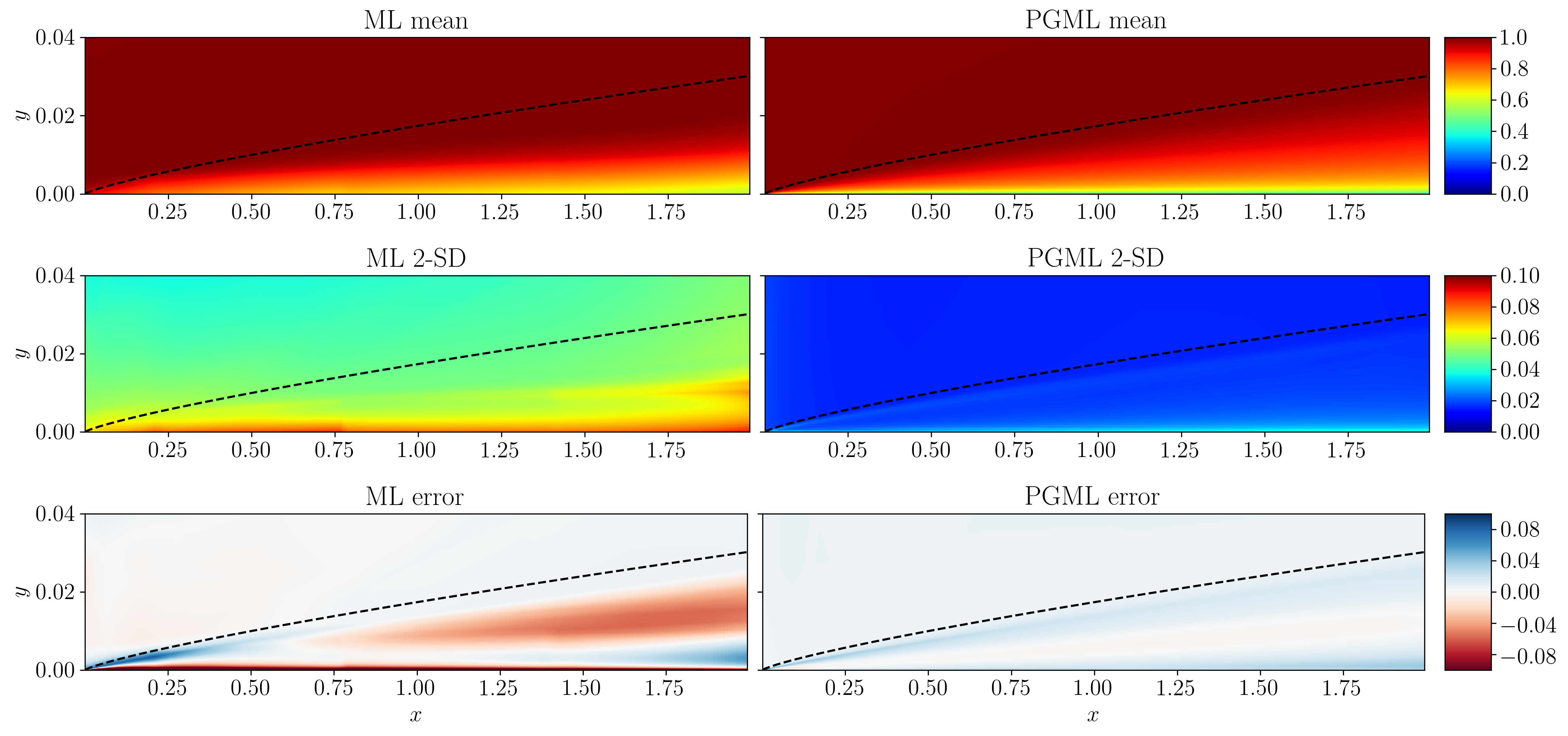}
\caption{Prediction of the turbulent flat plate boundary layer with 5\% of data used for training the ML and PGML model. The error is calculated as the difference between the true flow field and the flow field predicted by ML and PGML models.}
\label{fig:turbulent_0.05}
\end{figure*}

\begin{figure*}[htbp]
\centering
\includegraphics[width=0.99\textwidth]{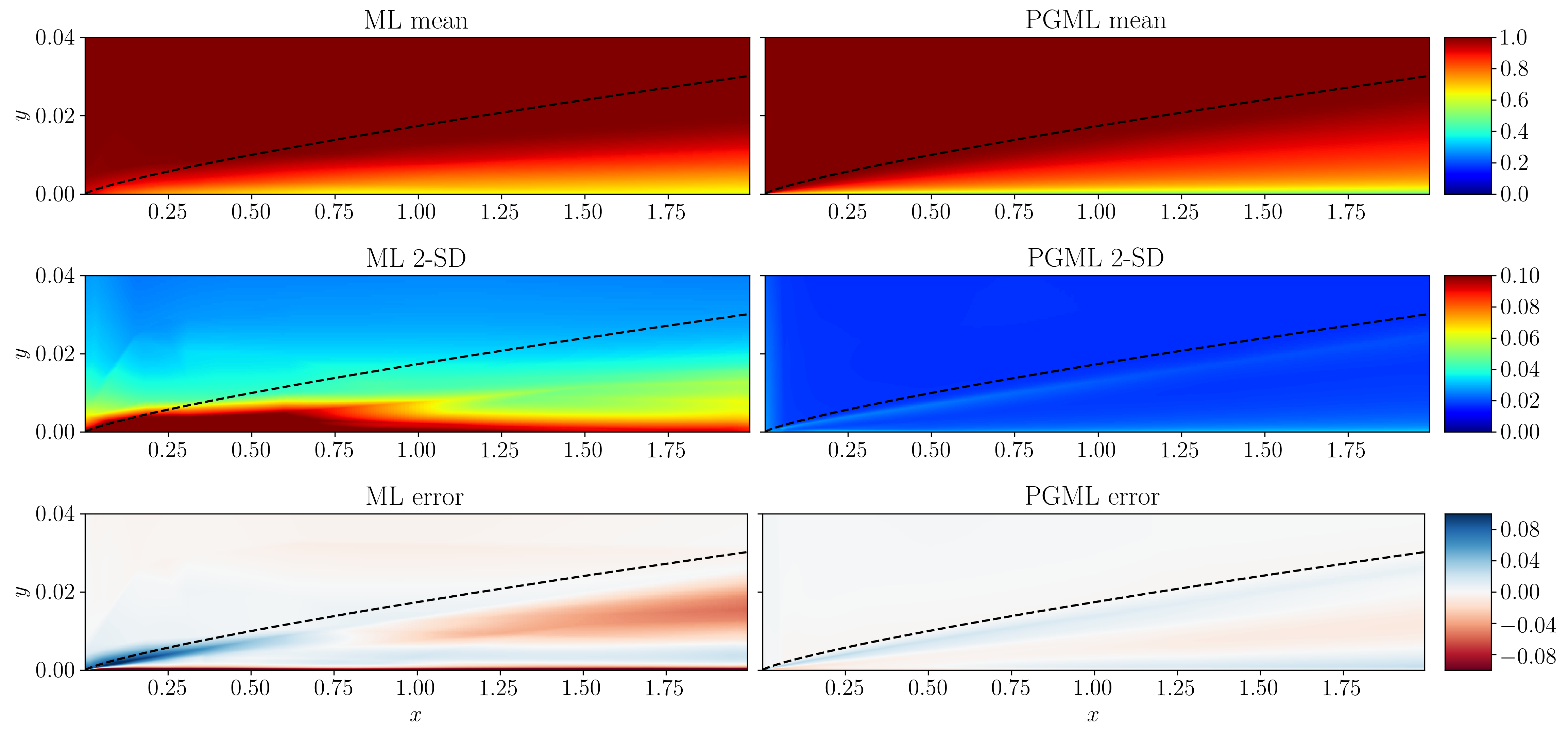}
\caption{Prediction of the turbulent flat plate boundary layer with 10\% of data used for training the ML and PGML model. The error is calculated as the difference between the true flow field and the flow field predicted by ML and PGML models.}
\label{fig:turbulent_0.1}
\end{figure*}

\begin{figure*}[htbp]
\centering
\includegraphics[width=0.99\textwidth]{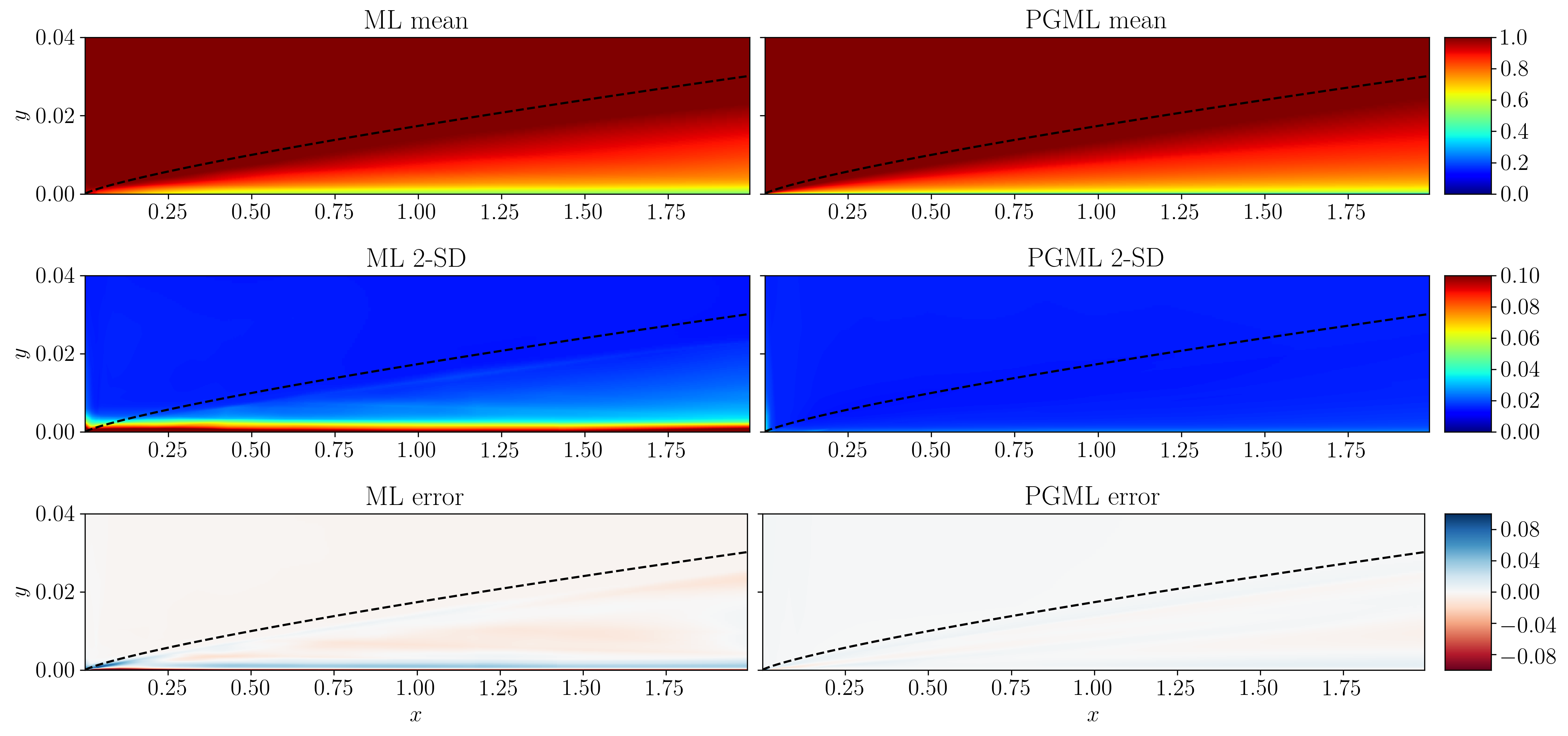}
\caption{Prediction of the turbulent flat plate boundary layer with 30\% of data used for training the ML and PGML model. The error is calculated as the difference between the true flow field and the flow field predicted by ML and PGML models.}
\label{fig:turbulent_0.3}
\end{figure*}

Figures~\ref{fig:turbulent_0.05}-~\ref{fig:turbulent_0.3} shows the spatial variation of the predicted mean of the velocity field, confidence interval of two standard deviations, and the error with respect to the true velocity field near the wall region for 5\%, 10\%, and 30\% of the training data, respectively. As the training data increases, the error decreases for both ML and PGML models.  The confidence estimate associated with the PGML model is substantially higher (i.e., lower uncertainty) than the ML model for all three datasets. Moreover, the error of the PGML model is greatly reduced compared to the ML model. One other benefit of constructing a PGML approach is its modular nature that can provide an opportunity of bridging the gap between domain-specific knowledge and physics-agnostic models.  

\begin{figure*}[htbp]
\centering
\includegraphics[width=0.99\textwidth]{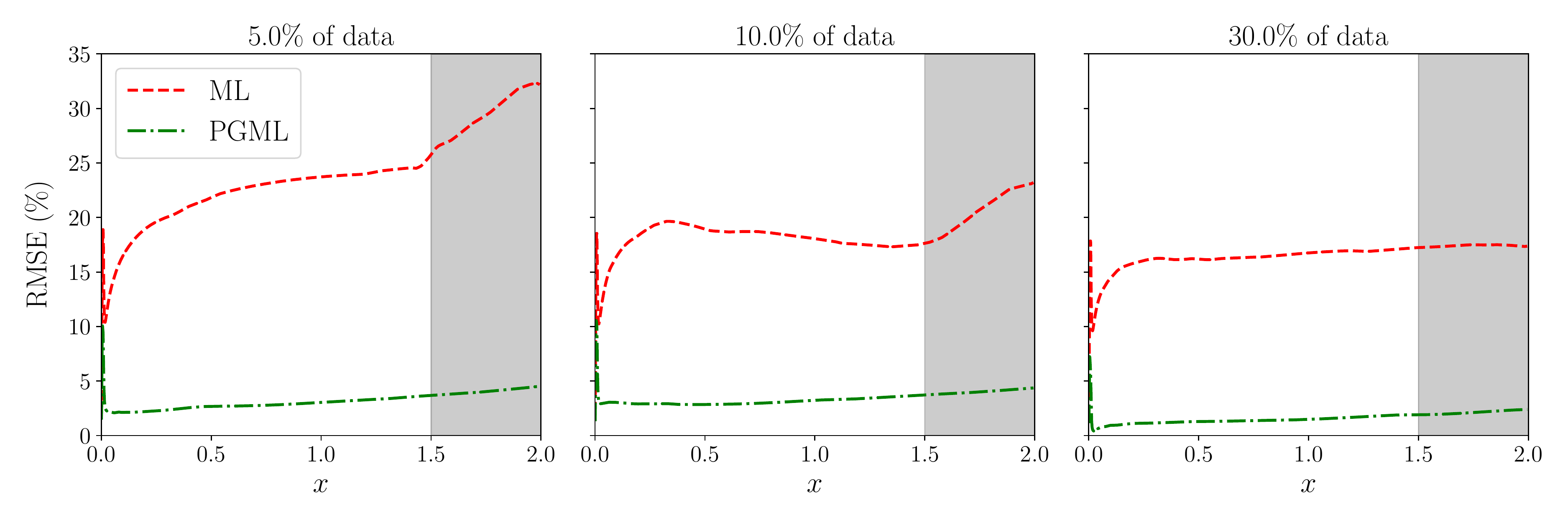}
\caption{Variation of the normalized RMSE (in percentage) along the streamwise direction for ML and PGML models. The amount of observations data used for training the model is 5\% (left), 10\% (middle), and 30\% (right). The gray color shaded area represents the extrapolation region. }
\label{fig:rmse}
\end{figure*}

In our previous numerical experiments, we focused on the reconstruction task within the interpolation region. Both ML and PGML models were trained using the data sampled from the whole domain, i.e., up to $L=2.0$, where $L$ is the length of the flat plate. In our next numerical experiments, we sample observations only from the region till $L=1.5$. Therefore, the region between $L=1.5$ to $L=2.0$ corresponds to the extrapolation region. We quantify the performance of the ML and PGML model using the variation of root mean squared error (RMSE) percentage along the streamwise direction as follows
\begin{equation}
    \text{RMSE}(x) = 100 \times \bigg(\frac{1}{N_y} \sum_{j=1}^{N_y} \bigg(\frac{u_T(y_j) - u_P(y_j)}{u_T(y_j)}\bigg)^2 \bigg)^{1/2}
\end{equation}
where $u_T$ is the velocity of the high-fidelity model, $u_p$ is the velocity predicted from the data-driven model, $N_y$ is the spatial resolution in the wall-normal direction. From Figure~\ref{fig:rmse}, we can see that the RMSE increases substantially in the extrapolation region for the ML model, especially when the observations are very sparse, i.e., 5\% of the data. This is a well-known limitation of DL models to extrapolate poorly in the absence of dense data. The PGML model on the other hand has RMSE almost one order of magnitude less than the ML model in the interpolation region. Additionally, the increase in RMSE is not significant in the extrapolation region. This demonstrates that the PGML model is robust for the unseen condition, and it performs well for out-of-distribution examples.

\section{Discussion}

This study aims to develop a physics-guided machine learning (PGML) framework to improve data-driven models using prior knowledge from low-fidelity models. The PGML is a new deep neural network architecture that makes it possible to inject known physics during the training and \emph{deployment} processes to reduce uncertainty and consequently improve the trade-off between efficiency and accuracy. Our design of a hierarchically sequential learning algorithm allows us to embed simplified theories, low order models, or empirical laws directly into deep learning models. These physics-based injections assist the neural network models in constraining the output to a manifold of the physically consistent solution and leads to improved reliability and generalizability. The PGML model trained using the deep ensembles algorithm provides us an estimate of the uncertainty associated with the prediction. This uncertainty information can be used for several applications like active learning, sensor placement, and optimization. Some of the questions addressed in this study are as follows
\begin{itemize}
    \item How prior information on the physics of the problem can be used to improve black-box machine learning models?
    \item Can a concatenated neural network architecture augmented by a simplified or empirical model outperform a pure data-driven reconstruction model? 
    \item How can these data-driven predictive models be quantified regarding their uncertainties?
    \item What is the generalizability of predictive performance across ML and PGML architectures, when these are applied to unseen conditions?
\end{itemize}

Our analysis indicates that an injection of the empirical relations like \emph{one-seventh power law} improves the predictive models significantly for estimating canonical turbulent flat plate boundary layer flows. We found that the PGML outperforms its ML counterpart by reducing the RMSE to nearly an order of magnitude lower levels. We also demonstrated that the proposed PGML framework substantially reduces the model uncertainty even when only sparse observations are available. Furthermore, generalizability of results is also supported when we integrate our predictive models for unseen conditions. Specifically, the RMSE distributions for the ML model (with respect to true data) are around 20\% and 30\% for interpolation and extrapolation regions, respectively. However, the PGML model has superior performance in both interpolation and extrapolation regions with the RMSE distribution in the range of 3\%. For laminar flows, we also showed that a PGML approach (via injecting the Blasius solution) outperforms the ML approach. The Blasius approximation can also be extended for heat transfer problems (based on Falkner-Skan solutions) and for strongly nonlinear problems using non-similarity solutions \cite{liao2009general}. The concatenated neural networks are capable of discovering a correlation between low- and high-fidelity data allowing for smaller training dataset sizes, training time, and improved extrapolation performance. Therefore, the PGML model has a great potential for a vast number of physical systems where a hierarchy of models is commonly used. We also highlight that the PGML approach could be useful for generating physics consistent initial conditions to accelerate large-scale high-fidelity computations with eliminating non-physical initial transient time.

\section{Methods}
\subsection{Multi-fidelity concatenated neural network}
A neural network is a computational graph composed of several layers consisting of the predefined number of neurons. Each neuron is associated with certain coefficients called weights and some bias. The input from the previous layer is multiplied by a weight matrix as shown below
\begin{equation}
    S^l = \mathbf{W}^l \mathbf{x}^{l-1},
\end{equation}
where $\mathbf{x}^{l-1}$ is the output of the $(l-1){\text{th}}$ layer, $\mathbf{W}^l$ is the matrix of weights for the $l{\text{th}}$ layer. The summation of the above input-weight product and the bias is then passed through a node's activation function which is usually some nonlinear function. The introduction of nonlinearity through activation function allows the neural network to learn highly complex relations between the input and output. The output of the $l{\text{th}}$ layer can be written as
\begin{equation}
    \mathbf{x}^l = \zeta(S^l+B^l),
\end{equation}
where $B^l$ is the vector of biasing parameters for the $l{\text{th}}$ layer and $\zeta$ is the activation function. If there are $L$ layers between the input and the output in a neural network, then the neural network mapping $\mathcal{F}:\mathbf{x}\rightarrow \mathbf{y}(\mathbf{x})$ can be represented as follows
\begin{equation}
    {\mathbf{y}} = \zeta_{L}(\cdot ; \boldsymbol{\Theta}_{L}) \circ \cdots \circ \zeta_{2}(\cdot ; \boldsymbol{\Theta}_{2}) \circ \zeta_{1}(\mathbf{x} ; \boldsymbol{\Theta}_{1})
\end{equation}
where $\boldsymbol{\Theta}$ represents the weight and bias of the corresponding layer of the neural network and $\mathbf{y}$ is the final output of the neural network. For the concatenated neural network, the information from the low-fidelity model is injected at a certain intermediate layer of the neural network as follows
\begin{equation} \label{eq:con}
    {\mathbf{y}} = \zeta_{L}(\cdot ; \boldsymbol{\Theta}_{L}) \circ \cdots 
    \circ \underbrace{\mathcal{C}(\zeta_{i}(\cdot ; \boldsymbol{\Theta}_{i}),h(\mathbf{x}))}_{\text{Concatenation layer}} \circ 
    \cdots \circ \zeta_{1}(\mathbf{x} ; \boldsymbol{\Theta}_{1}),
\end{equation}
where $\mathcal{C}(\cdot, \cdot)$ represents the concatenation operation and the information from the low-fidelity model, i.e., $h(\mathbf{x})$ is injected at $i$th layer, and $\boldsymbol{\Theta}$ are the trainable parameters of the corresponding layer. The concatenation operator takes the latent variables at a particular layer and combines them with information from the low-fidelity model to return a vector. Specifically, if the $i$th layer has $D_h$ neurons and $h(\mathbf{x})\in \mathbb{R}^{D_l}$, the output of the $i$th layer will be in $\mathbb{R}^{D_h+D_l}$. For the boundary layer reconstruction problem, we use a neural network with three hidden layers and twenty neurons in each hidden layer. The prediction from the low-fidelity model is concatenated at the second hidden layer. One important caveat in a concatenated neural network is the selection of an appropriate intermediate layer at which to inject the low-fidelity model information, and tools like AutoML can be applied to automate this search. The concatenation operator given by Equation~\ref{eq:con} can also be applied to all hidden layers simultaneously as
\begin{equation} \label{eq:con2}
    {\mathbf{y}} = \underbrace{ \mathcal{C}(\zeta_{L}(\cdot ; \boldsymbol{\Theta}_{L}),h(\mathbf{x})) \circ \cdots 
    \circ \mathcal{C}(\zeta_{i}(\cdot ; \boldsymbol{\Theta}_{i}),h(\mathbf{x})) \circ 
    \cdots \circ \mathcal{C}(\zeta_{1}(\cdot ; \boldsymbol{\Theta}_{1}),h(\mathbf{x})) }_{\text{Concatenation layers}}.
\end{equation}
We highlight here that the proposed PGML framework is modular and Equation~\ref{eq:con2} can be generalized for fusing information from multiple low-fidelity models. However, we dedicate this study to investigate the feasibility of the proposed PGML framework for only two levels of approximations. An additional optimization problem can be constructed to search the best architecture in terms of relevant hyperparameters such as the number of hidden layers, and the location and sparsity of the concatenation structure. Such auto PGML investigations will be a topic that we will pursue in our future works.

\subsection{Deep Ensembles: Training and prediction}
Deep learning algorithms like neural networks approximate the mapping from inputs to outputs using trainable parameters called weights and biases. The parameters of the neural network are determined through the minimization of the loss function. The prediction from the neural network is usually a point estimate, i.e., continuous outputs for regression tasks and discrete classes for classification problems. However, the information about the confidence in model's prediction might be crucial for many scientific applications \cite{amodei2016concrete}. The uncertainty estimates can also be useful for applications like sensor placement and Bayesian optimization. In this study, we apply the deep ensembles algorithm for estimating the probabilistic distribution function (PDF) of output conditioned on the inputs \cite{lakshminarayanan2017simple}. While there are state-of-the-art methods like Bayesian neural networks \cite{neal2012bayesian} that quantifies uncertainty by learning the distribution of weights, deep ensembles is adopted due to their simplicity and scalability.

Here, we briefly discuss the uncertainty quantification mechanism of deep ensembles. We assume that our training dataset $\mathcal{D}$ consists of $N$ samples $\mathcal{D}=\{\mathbf{x}_i, \mathbf{y}_i \}_{i=1}^{N}$, where $\mathbf{x} \in \mathbb{R}^P$ represents the $P-$dimensional features and the label is $Q-$dimensional, i.e., $\mathbf{y} \in \mathbb{R}^{Q}$. A neural network is trained by minimizing the loss function $L(\mathbf{y}, \tilde{\mathbf{y}}(\mathbf{x}; \theta))$, where $\tilde{\mathbf{y}}$ is the predicted label from a neural network parameterized by $\mathbf{\theta}$. The most common loss function for regression tasks is the mean squared error (MSE) between true and predicted labels averaged over all samples in the dataset. The MSE loss function does not give an estimate of the probability distribution of $p(\tilde{\mathbf{y}}|\mathbf{x})$ and hence the uncertainty estimate is usually absent with the prediction from neural networks. 

In order to quantify the predictive uncertainty, the neural network is trained to output the mean and variance of the Gaussian distribution in the output layer. The weights of the neural network are determined by minimizing the negative log-likelihood $\mathcal{L}$ as follows
\begin{equation}
    \mathbf{\theta} = \argmin_{\mathbf{\theta}}[\mathcal{L}] , \quad \text{where} \quad \mathcal{L} = \sum_{i=1}^N \frac{1}{2} \text{log} \sigma^2 (\mathbf{x}_i) + \frac{(\mathbf{y}_i - \mu(\mathbf{x}_i))^2}{2 \sigma^2 (\mathbf{x}_i)},
    \label{eq:nll}
\end{equation}
where the mean $\mu$ and the variance $\sigma^2$ are parameterized by the neural network. The positivity constraint is enforced for the variance by passing the output corresponding to variance of distribution though the \textit{softplus} function $\text{log}(1 + \text{exp}(\cdot))$, and adding a minimum variance (for example $10^{-6}$) for numerical stability. If we assume the variance to be \textit{constant} in Equation~\ref{eq:nll} (i.e., it does not depend on input features), then the negative log-likelihood loss function becomes analogous to the MSE loss function. Therefore, from a probabilistic point of view, minimizing the MSE is equivalent to minimizing negative log-likelihood with an assumption of Gaussian distribution with constant standard deviation \cite{davison2003statistical,nielsen2015neural}. 

The ensemble of neural networks has been demonstrated to be successful in improving the predictive performance of machine learning models \cite{dietterich2000ensemble}. There are broadly two methods of generating ensembles, (i) randomization-based approaches where the ensembles can be trained in parallel without any interaction, and (ii) boosting-based approaches where the ensembles are trained sequentially \cite{ferreira2012boosting}. The randomization procedure for generating ensembles of neural networks should be such that prediction from individual models are de-correlated and each individual models are strong (i.e., high accuracy). In this work, the random initialization of weights of the neural network is used for generating ensembles. There are other schemes such as bagging where the ensembles of neural networks are trained on a different subset of the original training data. However, random initialization is better than bagging for improving predictive accuracy and uncertainty \cite{lakshminarayanan2017simple,lee2015m}. This simple and yet robust randomization approach is highly scalable as it allows for distributed training of neural networks and can be applied to many scientific problems. For computing the predictive probability distribution, we approximate the ensemble prediction as a Gaussian whose mean and variance are computed as follows
\begin{align}
    \mu_{*}(\mathbf{x}) &= \frac{1}{M}\sum_{j=1}^M \mu_{\mathbf{\theta}_j}(\mathbf{x}), \\
    \sigma_{*}^2(\mathbf{x}) &= \frac{1}{M}\sum_{j=1}^M (\sigma_{\mathbf{\theta}_j}^2(\mathbf{x}) +  \mu_{\mathbf{\theta}_j}^2(\mathbf{x}) ) - \mu_{*}^2(\mathbf{x}),
\end{align}
where $\mu_{\mathbf{\theta}_j}$ and $\sigma_{\mathbf{\theta}_j}$ is the mean and standard deviation of predicted probability distribution by the $j$th neural network. We employ an ensemble of five neural networks in this study (i.e., $M=5$).  



\subsection{Multi-fidelity data fusion for laminar boundary layer}
The first test case considered in this study is the laminar boundary layer flow. Boundary layer flows can be characterized by dividing the flow into two regions, one inside the boundary layer where the viscosity dominates and one outside the boundary layer where the effect of viscosity can be neglected. The low-fidelity model considered for laminar flow is the steady-state two-dimensional laminar boundary layer described using Blasius equation \cite{white2006viscous}. The core idea behind Blasius equation is transforming a partial differential equation (PDE) comprised of the flat plate boundary layer equations, with zero pressure gradient, into a single ordinary differential equation (ODE) by using a similarity solution approach. The derivation of the Blasius equation can be found in many texts on fluid mechanics and we describe only the final form. The Blasius equation and its boundary conditions can be written as 
\begin{align}
    2f{'''} + ff'' = 0, \label{eq:blasius}\\
    f(0) = f'(0) = 0, \  f'(\infty) = 1, \label{eq:blasius2}
\end{align}
where $f(\eta)$ is a function of similarity variable $\eta$. The similarity variable $\eta$ is defined as $\eta = y\sqrt{u_\infty/{(x\nu)}}$, where $y$ is the direction normal to the plate, $x$ is the direction along its length with zero being the leading edge, $u_\infty$ is the freestream velocity, and $\nu$ is the kinematic viscosity of the fluid. The third-order ODE is first split into a coupled system of three first-order ODEs. Then we apply the shooting method to determine the initial value for $f''(0)$, and the first-order ODEs are numerically integrated with the fourth-order Runge-Kutta scheme \cite{moin2010fundamentals}. The velocity profile from the Blasius solution can be determined using the relation $\bar{u}=u_\infty f'$, where overbar symbol is used to indicate the low-fidelity model estimate. The high-fidelity observations are generated by solving the incompressible Navier-Stokes equations with the PISO algorithm \cite{issa1986solution} available in OpenFoam. We get the velocity (components along streamwise and wall-normal directions) and the pressure distribution from CFD simulation. The Reynolds number based on the length of the flat plate used for generating data is $\text{Re}_L=5 \times 10^{4}$, where $L$ is the length of the flat plate. The training data for the concatenated neural network is sampled randomly from the whole domain and the velocity field is contaminated by adding a white Gaussian noise with zero mean and a standard deviation of 0.05. Advanced sampling methods like Latin hypercube sampling, clustered sampling can be utilized to reduce the number of samples required for training and we will consider this as part of our future work.

\begin{figure*}[t!]
\centering
\includegraphics[width=0.99\textwidth]{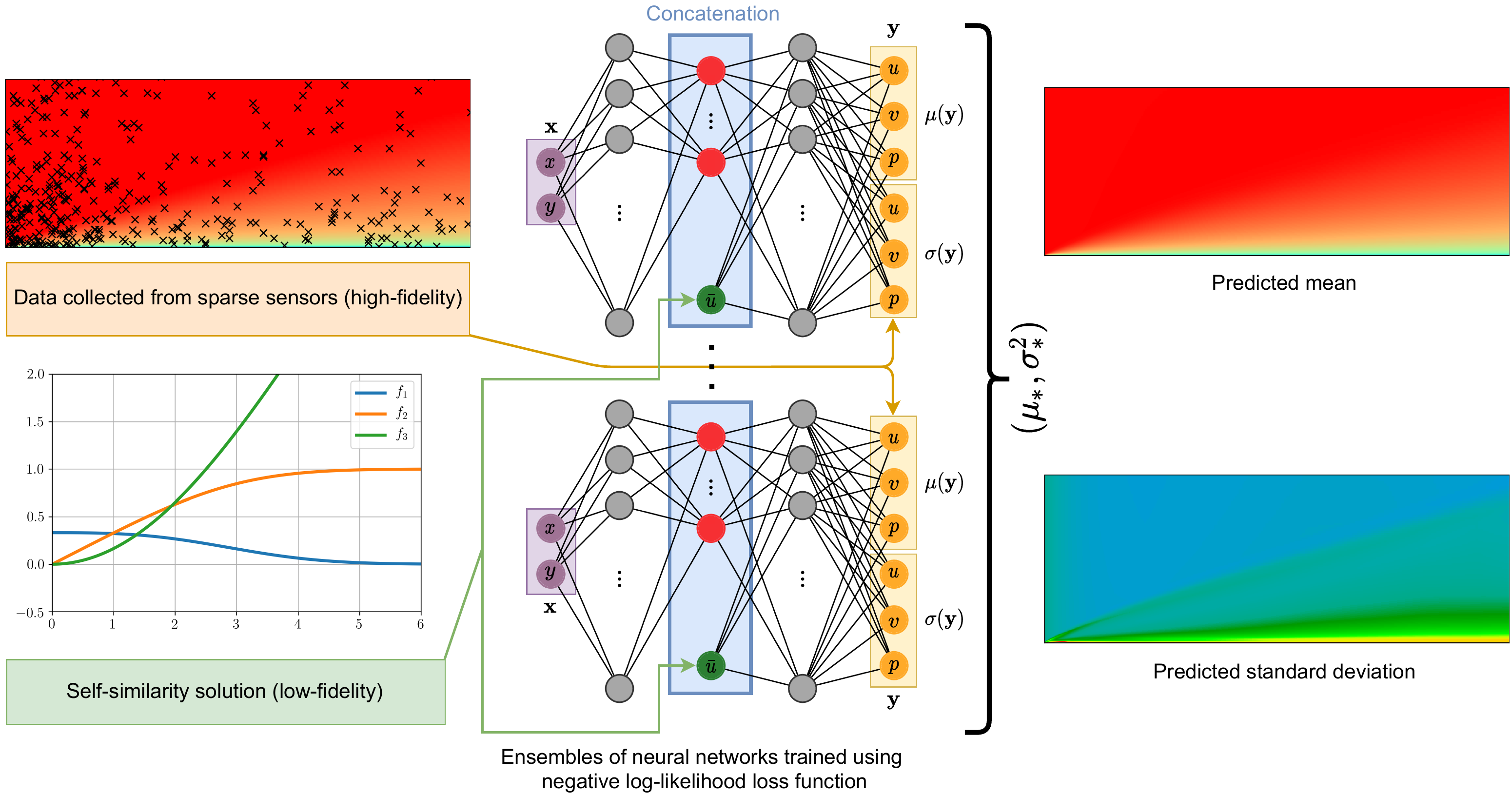}
\caption{Illustration of the multi-fidelity data-fusion framework applied to the flat plate boundary layer prediction task. The self-similarity Blasius solution is replaced by the one-seventh power law when applying to the turbulent boundary layer flows. 
}
\label{fig:pgml2}
\end{figure*}

For the laminar boundary layer reconstruction task, the input to the neural network is the location of the sensor, i.e., $\mathbf{x}=[x,y]$, where $x$ and $y$ are the positions of the sensors in streamwise and wall-normal directions. The output of the neural network is the probability distribution of $u$, $v$, $p$ represented by their mean and standard deviation, where $u$ is the velocity in the streamwise direction, $v$ is the velocity in the wall-normal direction, and $p$ is the pressure at the sensor's location. Additionally, the velocity profile obtained from the Blasius solution is used as the low-fidelity model, i.e., $h(\mathbf{x})=[\bar{u}]$. Following our previous discussion, the problem formulation can be written as 
\begin{equation}
    \{\mu(\mathbf{x}), \sigma(\mathbf{x}) \} = \mathcal{F}(\mathbf{x}, h(\mathbf{x})), \quad p(\mathbf{y}|\mathbf{x}, h(\mathbf{x})) = \mathcal{N}(\mu_{*}(\mathbf{x}), \sigma_{*}(\mathbf{x})).
    \label{eq:laminar_problem}
\end{equation}
In a nutshell, Figure~\ref{fig:pgml2} shows the proposed multi-fidelity data-fusion framework for laminar boundary layer flows. 

\subsection{Multi-fidelity data fusion for turbulent boundary layer}
The boundary layer around the flat plate transitions to turbulence at high Reynolds number. Before the advent of supercomputing, it was not possible to numerically solve the Navier-Stokes equations for turbulent flows and fluid dynamicists had to resort to experimental studies to derive empirical relations for high Reynolds number flows. In this study, the low-fidelity approximation for the turbulent boundary layer is obtained using the one-seventh power law. The one-seventh power law \cite{yunus2010fluid} for computing the mean (or ensemble-averaged) velocity profile for flat-plate turbulent boundary layer is given as follows 
\begin{equation}
\frac{\bar{u}}{u_\infty} \approx \begin{cases}
  \big( \frac{y}{\delta} \big)^{1/7} \quad &\text{for} \quad y \le \delta, \\
  1 \quad &\text{for} \quad y > \delta,
\end{cases}
\label{eq:power_law}
\end{equation}
where $u_\infty$ is the freestream velocity, $y$ is the direction normal to the plate, and the turbulent boundary layer thickness $\delta$ is computed as follows \cite{yunus2010fluid}
\begin{equation}
    \delta \approx \frac{0.38 x}{(\text{Re}_x)^{1/5}}, 
\end{equation}
where $\text{Re}_x$ is the Reynolds number at a given $x$-location. There are many such empirical relations available to approximate turbulent boundary layers such as the log law, and Spalding's law of the wall \cite{spalding1961single}. The high-fidelity data is generated for the flat-plate turbulent boundary layer by solving the incompressible Navier-Stokes equations with the SIMPLE algorithm and $k-\omega$-SST turbulent model implemented in OpenFoam. The Reynolds number based on the length of the flat plate for turbulent boundary layer simulation is $\text{Re}_L=1 \times 10^{7}$. One important parameter in turbulence modeling is the dimensionless distance in the normal direction called wall $y^+$ and is defined as $y^+=\sqrt{y u_\tau/\nu}$, where $u_\tau$ is the friction velocity. The friction velocity is calculated based on the wall shear stress as $u_\tau = \sqrt{\tau_w/\rho}$. The mesh is refined near the flat plate in such a way that the near-wall $y^+$ is below 5. The locations for collecting the data are sampled randomly and are contaminated by adding a white Gaussian noise with zero mean and a standard deviation of 1.0 to mimic the measurement error. The formulation of turbulent boundary layer reconstruction is similar to the laminar boundary reconstruction as given in Equation~\ref{eq:laminar_problem} except for the low-fidelity model. The low-fidelity model prediction for turbulent boundary layer flow is calculated using Equation~\ref{eq:power_law}.

\section*{Acknowledgements}
This material is based upon work supported by the U.S. Department of Energy, Office of Science, Office of Advanced Scientific Computing Research under Award Number DE-SC0019290. O.S. gratefully acknowledges their support. 
Disclaimer: This report was prepared as an account of work sponsored by an agency of the United States Government. Neither the United States Government nor any agency thereof, nor any of their employees, makes any warranty, express or implied, or assumes any legal liability or responsibility for the accuracy, completeness, or usefulness of any information, apparatus, product, or process disclosed, or represents that its use would not infringe privately owned rights. Reference herein to any specific commercial product, process, or service by trade name, trademark, manufacturer, or otherwise does not necessarily constitute or imply its endorsement, recommendation, or favoring by the United States Government or any agency thereof. The views and opinions of authors expressed herein do not necessarily state or reflect those of the United States Government or any agency thereof.

\appendix


\bibliographystyle{unsrt} 
\bibliography{ref}

\end{document}